\documentclass[runningheads,a4paper]{llncs}

\usepackage{llncsdoc}
\usepackage{subfigure}
\usepackage{epsfig}
\usepackage{fancyhdr}
\usepackage{color}
\usepackage{multirow}
\usepackage{graphicx}
\usepackage{float}

\usepackage{threeparttable}
\usepackage{hyperref}
\hypersetup{breaklinks=true}
\usepackage[hyphenbreaks]{breakurl}

\begin{document}

\title{BigDataBench-MT: A Benchmark Tool for Generating Realistic Mixed Data Center Workloads}
\author{$^{1}$Rui Han, $^{3}$Shulin Zhan, $^{4}$Chenrong Shao,  $^{5}$Junwei Wang, \\
         $^{6}$Lizy K. John, $^{1}$Jiangtao Xu, $^{1,2}$Gang Lu, $^{1}$Lei Wang}
\institute{$^{1}$Institute Of Computing Technology, Chinese Academy of Sciences, Beijing, China\\
    $^{2}$University of Chinese Academy of Sciences, China\\
    $^{3}$iCarsclub, Beijing, China\\
    $^4$Xi'an Jiaotong University, Xi'an, China\\
    $^5$Kingsoft Cloud, Beijing, China\\
    $^{6}$Department of Electrical and Computer Engineering, the University of Texas at Austin, TX, USA\\
\email{hanrui@ict.ac.cn, zhanshulin@ppzuche.com, scr1994@stu.xjtu.edu.cn, wangjunwei@kingsoft.com, ljohn@ece.utexas.edu, \{xujiangtao, lugang, wl\}@ict.ac.cn}
}
\maketitle
%, $^{1,2}$Jianfeng Zhan

\begin{abstract}
Long-running service workloads (e.g. web search engine) and short-term data analysis workloads (e.g. Hadoop MapReduce jobs) co-locate in today's data centers. Developing realistic benchmarks to reflect such practical scenario of mixed workload is a key problem to produce trustworthy results when evaluating and comparing data center systems. This requires using actual workloads as well as guaranteeing their submissions to follow patterns hidden in real-world traces.
However, existing benchmarks either generate actual workloads based on probability models, or replay real-world workload traces using basic I/O operations.
To fill this gap, we propose a benchmark tool that is a first step towards generating a mix of actual service and data analysis workloads on the basis of real workload traces. Our tool includes a combiner that enables the replaying of actual workloads according to the workload traces, and a multi-tenant generator that flexibly scales the workloads up and down according to users' requirements. Based on this, our demo illustrates the workload customization and generation process using a visual interface.
The proposed tool, called BigDataBench-MT, is a multi-tenant version of our comprehensive benchmark suite BigDataBench and it is publicly available from \url{http://prof.ict.ac.cn/BigDataBench/multi-tenancyversion/}.
\end{abstract}

% Considering our community may feel interest in using these workloads to evaluate new system designs and implementations, our tool and the corresponding workload traces are publicly available from http://prof.ict.ac.cn/BigDataBench/multi-tenancyversion/.


\section{Introduction}  % Article style
In modern cloud data centers, a large number of tenants are consolidated to share a common computing infrastructure and execute a diverse mix of workloads. Benchmarking and understanding these workloads is a key problem for system designers, programmers and researchers to optimize the performance and energy efficiency of data center systems and to promote the development of data center technology. This work focuses on two classes of popular data center workloads \cite{reiss2012heterogeneity}:
\begin{itemize}
\item \emph{Long-running services}. These workloads offer online services such as web search engines and e-commerce sites to end users and the services usually keep running for months and years. The \emph{tenants} of such workloads are \emph{service end users}.
 \item \emph{Short-term data analysis jobs}. These workloads process input data of many scales using relatively short periods (e.g. in Google and Facebook data centers, a majority (over 90\%) of analytic jobs complete within a few minutes \cite{chen2012interactive,reiss2012heterogeneity}). The \emph{tenants} of such workloads are \emph{job submitters}.
\end{itemize}

%we believe it will be of interest to the data management community and a large user base to benchmark and understand the mix of these workloads.

As data analysis systems such as Hadoop and Spark mature, both types of workloads widely co-locate in today's data centers, hence the pressure to benchmark and understand these mixed workloads rises. Within this context, we believe that it will be of interest to the data management community and a large user base to generate realistic workloads such that trustworthy benchmarking reflecting the practical data center scenarios can be conducted. Considering the heterogeneity and dynamic nature of data center workloads and their aggregated resource demands and arrival patterns, this requires overcoming two major challenges.

\textbf{Benchmarking using actual workloads based on real-world workload traces}. Data analysis jobs usually have various computation semantics (i.e. implementation logics or source codes) and input data sizes (e.g. ranging from KB to PB), and their behaviors also heavily rely on the underlying software stacks (such as Hadoop or MPI). Hence it is difficult to emulate the behaviors of such highly diverse workloads just using synthetic workloads such as I/O operations. On the other hand, generating workloads whose arrival patterns follow real-world traces is an equally important aspect of realistic workloads.
This is because these traces are the most realistic data sources including both explicit and implicit arrival patterns (e.g. sequences of time stamped requests or jobs).

\textbf{Benchmarking using scalable workloads with realistic mixes}. A good benchmark needs to flexibly adjust the scale of workloads to meet the requirements of different benchmarking scenarios. Based on our experience, we noticed that in many cases, obtaining real workload traces is difficult due to confidential issues. The limited trace data also restrict the scalability of benchmark. It is therefore challenging to produce workloads at different scales while still guaranteeing their realistic mix corresponding to real-world scenarios.

% Existing benchmarks
% Realistic data source as they include all explicit and implicit job patterns
% and YCSB \cite{cooper2010benchmarking}

%OUR WORK
In this paper, we propose a benchmark tool that is a first step towards generating realistic mixed data center workloads. This tool, called BigDataBench-MT, is a multi-tenancy version of our open-source project BigDataBench, which is a comprehensive benchmark suite including 14 real-world data sets and 33 actual workloads covering five application domains \cite{opensourceBigDataBench,wang2014bigdatabench}. The goal of BigDataBench-MT is not only supporting the generation of service and data analysis workloads based on real workload traces, but also providing a multi-tenant framework to enable the scaling up and down of such workloads with guarantee of their realistic mixes. Considering our community may feel interest in using these workloads to evaluate new system designs and implementations, our tool and the corresponding workload traces are publicly available from \url{http://prof.ict.ac.cn/BigDataBench/multi-tenancyversion/}.

\section{Related Work}

We now review existing data center benchmarks from three perspectives, as shown in Table \ref{tab:applications}.
%\begin{itemize}
%\item

\textbf{Evaluated platform}. First of all, we classify data center benchmarks according to their targeted systems. We consider three popular camps of systems in today's data centers: (1) \emph{Hadoop-related systems}: the great prosperity of the Hadoop-centric systems in industry brings a wide diversity of systems (e.g. Spark \cite{Spark}, HBase\cite{HBase2015}, Hive \cite{Hive} and Impala \cite{Impala}) on top of Hadoop MapReduce and HDFS \cite{HadoopEcosystems} as well as a wide range of benchmarks specifically designed for these systems. (2) \emph{Data stores}: parallel DBMSs (e.g. MySQL \cite{mysql} and Oracle \cite{oracle}) and NoSQL data stores (e.g. Amazon Dynamo \cite{decandia2007dynamo}, Cassandra \cite{lakshman2010cassandra} and Linkedin Voldemort \cite{sumbaly2012serving}) also widely exist in data centers. (3) \emph{Web services}: long-running web services such as Nutch search engine \cite{nutchsearch} and multi-tier cloud applications \cite{han2014enabling} are another important type of data center applications. These services usually have stringent response time requirement \cite{han2015sarp} and their request processing is distributed into a large number of service components for parallel processing, thus the service latency is determined by the tail latency of these components \cite{dean2013tail,han2015pcs}.

\textbf{Workload implementation logic}. Consider the complexity and diversity of workload behaviors in current data center systems, the implementation logic of existing data center benchmarks can be classified into three categories. The first category of benchmarks implement their workloads with algorithms. For example, HiBench \cite{huang2010hibench} include workloads implemented with machine learning algorithms in Mahout \cite{mahout}. The second category of benchmarks implement workloads using database operations such as reading, loading, joining, grouping, unifying, ordering, aggregating and spliting data. The third category of benchmarks implement workloads as I/O operations. For example, NNBench \cite{NNBench} and TestDFSIO \cite{DFSIO} emulate I/O operations on Hadoop HDFS; GridMix \cite{gridmix} provides two workloads: LoadJob that performs I/O operations and SleepJob that sleeps the jobs; and SWIM \cite{chen2012interactive} provides four workloads that stimulate the operations of Hadoop jobs to read, write, shuffle and sort data.
We view the first two categories of workloads as \emph{actual workloads}, because these workloads have semantics and they consume resources of processors, memories, caches and I/O bandwidths in execution. By contrast, workloads belonging to third category only consume I/O resources.

\textbf{Workload mix}. Finally, we classify data center benchmarks into three categories from the perspective of workload mix. The first type of data center benchmarks either generate single workloads (e.g. WordCount~\cite{wordcount}, Grep~\cite{grep} and Sort~\cite{sort}) or generate multiple workloads individually (e.g. CALDA~\cite{pavlo2009comparison}, AMPLab benchmark~\cite{AMPBenchmark} and CloudSuite~\cite{ferdman2011clearing}). That is, these benchmarks donot consider workload mix.
The second category of benchmarks generate synthetic mixes of workloads. Many benchmarks (e.g. PigMix \cite{PigMix2013}, HcBench \cite{saletore2013hcbench} and BigBench \cite{ghazal2013bigbench}) generate mixes of workloads by manually determining their proportions. Similarly, TPC benchmarks \cite{tpc} design a query set as a synthetic mix of queries with different proportions. YCSB \cite{cooper2010benchmarking} uses a package to include a set of related workloads.
MRBS decides the frequencies of different workloads using probability distributions such as a random distribution.
Finally, third category of benchmarks generate a realistic mix of synthetic workloads whose arrival patterns faithfully follow real-world traces. For example, GridMix \cite{gridmix} and SWIM \cite{swim,chen2012interactive} first build a job trace to describe the realistic job mix by mining production loads, and then run synthetic I/O operations according to the trace. However, how to generate actual workloads on the basis of real workload traces is still an open question.

%Both
%based on a production cluster such as the Rumen trace, then runs a realistic mix of synthetic jobs (either LOADJOB or SLEEPJOB jobs) according to the trace.

%In addition to database benchmarks such as the TPC series of benchmarks, existing benchmarks designed for batch or analytical systems (e.g. Hadoop, Spark and Hive) can be divided into two complementary categories \cite{han2014big}.

%The first category of benchmarks \cite{AMPBenchmark,ampbig,ghazal2013bigbench,armstrong2013linkbench} supports actual workloads but the arrival patterns of these workloads are determined by probability models.

%By contrast, in the second category of benchmarks, the workload arrival patterns faithfully follow real workload traces, but only synthetic workloads are supported.

%Specifically, GridMix \cite{gridmix} provides two synthetic workloads: LoadJob that performs I/O operations and SleepJob that sleeps the jobs. SWIM \cite{chen2012interactive} provides four synthetic workloads that emulate the operations of Hadoop jobs to read, write, shuffle and sort data.

%\emph{Table: all existing benchmarks in the area of big data?}

%\emph{We consider four types of benchmarks, discuss}

%\textbf{Three dimensionality}
%+
%\textbf{Some benchmarks have overlaps, because they include multiple workloads}
%+
%Focus on big data workloads --> data analysis
%Service workload is straight-forward

\begin{table*} [!t]
\caption{Overview of data center benchmarks}
\centering
\begin{threeparttable}
\begin{tabular}{|p{1.6cm}|p{3.6cm}|p{4.2cm}|p{2.4cm}|}
\hline
\multirow{2}{1.6cm}{\textbf{Workload mix}} & \multicolumn{3}{|c|}{\textbf{Workload implementation logic}}\\
\cline{2-4}
&\textbf{Algorithms} & \textbf{Database operations} & \textbf{I/O operations}\\
\hline
\textbf{No mix}&WordCount$^1$\cite{wordcount}, Grep$^1$\cite{grep}, Sort$^1$\cite{sort}, Terasort$^1$\cite{terasort}, HiBench$^1$ \cite{huang2010hibench}, TPCx-HS$^1$\cite{nambiar2014standard}, Graphalytics$^1$\cite{capotua2015graphalytics}, CloudSuite$^4$\cite{ferdman2011clearing}
&MRBench$^1$\cite{kim2008mrbench}, CALDA$^2$\cite{pavlo2009comparison}, AMPLab benchmark$^2$\cite{AMPBenchmark}, YCSB$^2$\cite{cooper2010benchmarking}, BG benchmark$^2$\cite{barahmand2013bg}, CloudSuite$^4$\cite{ferdman2011clearing}
&NNBench$^1$\cite{NNBench}, TestDFSIO$^1$\cite{DFSIO}, HiBD$^1$\cite{shankar2014micro}\\
\hline

\textbf{Synthetic mix}& HcBench$^1$\cite{saletore2013hcbench}, MRBS$^1$\cite{sangroya2013mrbs}
&PigMix$^1$\cite{PigMix2013}, HcBench$^1$\cite{saletore2013hcbench}, MRBS$^1$\cite{sangroya2013mrbs}, BigBench$^2$\cite{ghazal2013bigbench}, LinkBench$^2$\cite{armstrong2013linkbench}, TPC benchmarks$^2$\cite{tpc}, TPC-W$^3$\cite{tpcw}, BigDataBench$^4$\cite{wang2014bigdatabench}
&HiBench$^1$\cite{huang2010hibench}, SPECWeb99$^3$\cite{specweb}\\
\hline
\textbf{Realistic mix}&&&Gridmix$^1$\cite{gridmix}, SWIM$^1$\cite{swim,chen2012interactive}\\
%BigDataBench$^3$\cite{wang2014bigdatabench}
\hline
\end{tabular}
\begin{tablenotes}
\item[$^1$] Hadoop-related systems
\item[$^2$] Data stores
\item[$^3$] Web services
\item[$^4$] All three types of systems
\end{tablenotes}
\label{tab:applications}
\end{threeparttable}
\end{table*}

\section{System Overview}

\begin{figure}
\centering
  \includegraphics[scale=0.55]{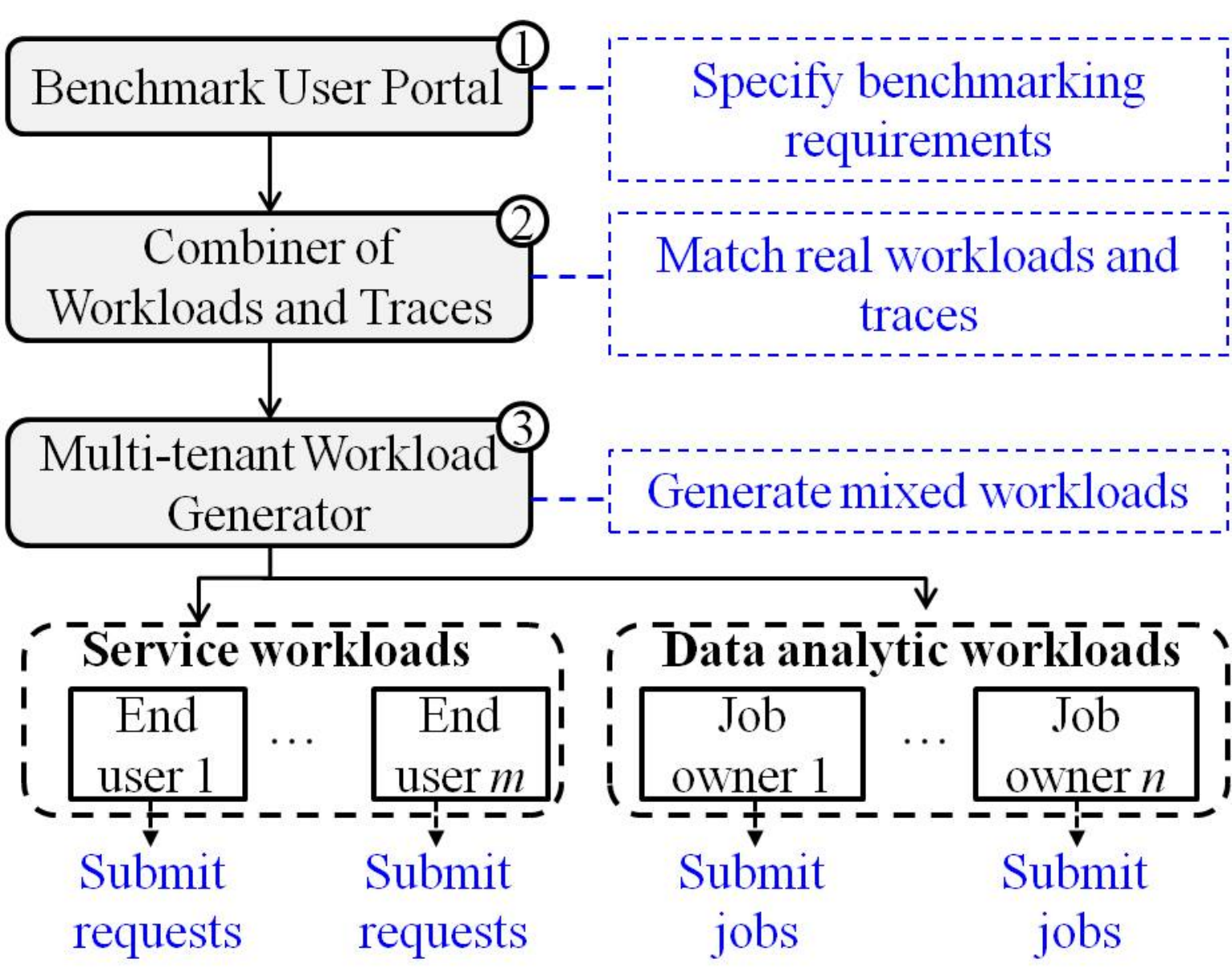}\\
  \caption{The BigDataBench-MT framework}
  \label{Fig: Overview}
\end{figure}

Figure \ref{Fig: Overview} shows the framework of our benchmark tool. It consists of three main modules. In the \emph{Benchmark User Portal}, users can first specify their benchmarking requirements, including the machine type and number to be tested, and the types of workload to use. A set of workload traces following these requirements are then selected. The next step of \emph{Combiner of Workloads and Traces} is to match the real workload and the selected workload traces, and outputs workload replaying scripts to guide the workload generation. Finally, the \emph{Multi-tenant Workload Generator} extracts the tenant information from the scripts and constructs a multi-tenant framework to generate a mix of service and data analysis workloads.

In BigDataBench-MT, we employ the Sogou user query logs \cite{sogoulogs} as the basis to generate the service workload (i.e. the Nutch search engine \cite{nutchsearch}) and the Google cluster workload trace as the basis to generate data analysis workloads (i.e. Hadoop and Shark workloads). The Sogou trace records logs from 50 days and it includes over 9 million users and 43 million queries. The Google trace records logs from 29 days and 12,492 machines and it includes over 5K users, 40K workload types, 1000K jobs and 144 million tasks. As a preprocessing step, we converted both traces into Impala databases (full version) and MySQL database (24-hour version) to facilitate the customization of benchmarking scenarios.
In the following subsections, we describe the last two modules of our tool.
%: Impala databases \cite{impalaWebsite} to store the complete version of workload traces and the MySQL databases \cite{mysqlWebsite} to store a one-day version for demonstration.

%\subsection{Benchmarking methodology} \label{Section: Benchmarking methodology}

%Facebook production workflow

\subsection{Combiner of Workload and Traces}  \label{Section: Combiner of Workload and Workload Traces}

%In contrast to conventional way of generating actual workloads whose arrival patterns are determined by probability models,

The goal of the combiner is to extract the request/job arrival patterns from real-world traces and combine them with actual workloads. The combiner applies differentiated combination techniques to the service and data analysis workloads because their workload generations have different features.

%The basic idea,
%submission patterns--> real-world scenarios

\textbf{Service workloads}. The generation of a service workload is determined by three factors: the request submitting time, the sequence of requests and the content of each request query. Take the web search engine for example, the combiner implements a request submitting service that automatically derives these factors from the Sogou trace and uses them to determine the request submission process.

\textbf{Data analysis workloads}. The generation of a data analysis workload is determined by four factors: the job submitting time, the workload type (i.e. the computation semantics and the software stack) and the input data (i.e. the data source and size). The current workload traces usually show the information of job submitting time but only provide \emph{anonymous jobs} whose workload types and/or input data are unknown. Hence the basic idea of the combiner is to derive the workload characteristics of both actual and anonymous jobs and then match jobs whose workload characteristics are sufficiently similar. Table \ref{table: metrics} lists the metrics used to represent workload characteristics, which reflect both jobs' performance (execution time and resource usage) and micro-architectural behaviors (CPI and MAI).

%Based the these traces, the existing benchmarks generate synthetic workloads

\begin{table}[h!]
  \caption{Metrics to represent workload characteristics of data analysis jobs}
  \centering
  \begin{tabular}{|c|l|}
    \hline
    \textbf{Metric} & \textbf{Description} \\
    \hline
    Execution time  & Measured in seconds\\
    \hline
    CPU usage   & Total CPU time per second\\
    \hline
    Total memory size   & Measured in GB\\
    \hline
    CPI   & Cycles per instruction\\
    \hline
    MAI & The number of memory accesses \\
    & per instruction\\
    \hline
  \end{tabular}
  \label{table: metrics}
\end{table}

\begin{figure}
\centering
  \includegraphics[scale=0.60]{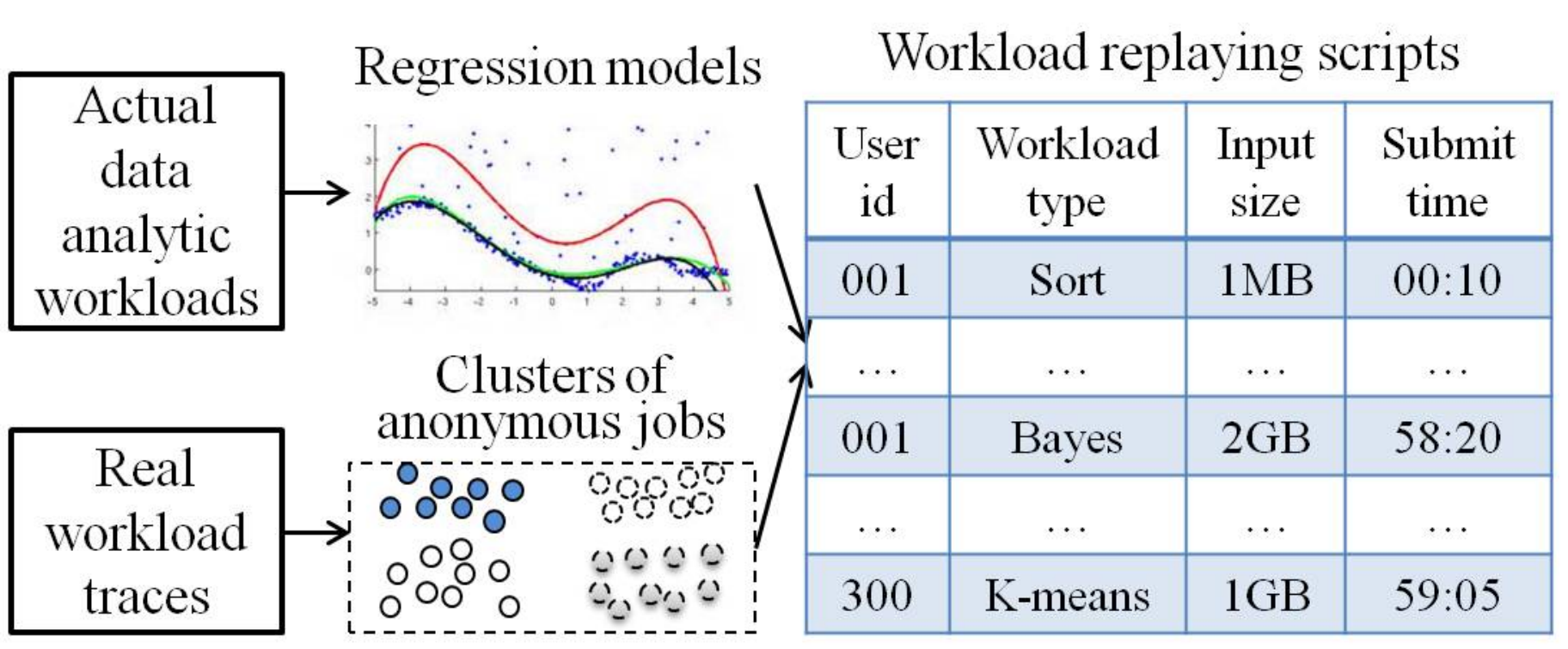}\\
  \caption{The matching process of real and synthetic data analysis jobs}
  \label{Fig: Combination}
\end{figure}

Figure \ref{Fig: Combination} shows the process of matching actual data analysis jobs and traces' anonymous jobs and it consists of two parallel sub-processes.
First, the actual jobs with different input data sizes are tested and their metrics of workload characteristics are collected. In BigDataBench-MT, we provide auto-running scripts to collect performance metrics and hardware performance counters (Perf \cite{PerfWebsite} and Oprofile \cite{OprofileWebsite} for Linux 2.6+ based systems) to obtain micro-architectural metrics.
Using the testing results as samples, the combiner trains the multivariate regression model to describe the relationship between an actual job (including both its workload type and input size as the independent variables) and its workload characteristic metrics (one metric is a dependent variable).
Second, the combiner views each anonymous job as an entity and the five workload characteristic metrics as its attributes, and employs the Bayesian Information Criterion (BIC)-based k-means clustering algorithm \cite{pelleg2000x} to group anonymous jobs in the trace into different clusters.

Based on the constructed regression models and clusters, the combiner further matches each cluster to one actual job with a specific input data. In the matching, the coefficient of variation (CV) measure, defined as the ratio of the standard deviation $\sigma$ to the mean $\mu$, is used to describe the dispersion of jobs in the same cluster. The combiner iteratively tests actual jobs of different workload types and input sizes, and matches an actual job with a cluster under two conditions: (i) the CV of the cluster is smaller than a specified threshold (e.g. 0.5), which indicates the anonymous jobs in this cluster are closely similar to each other; (ii) the change in this CV is smaller than a threshold (e.g. 0.1) after the actual job is added to the cluster. This means the workload characteristics of the added job are sufficiently similar to those of the anonymous jobs in the cluster. If multiple matched actual jobs are found for one cluster, the combiner selects the job resulting the smallest CV change. Finally, the combiner produces workload replaying scripts as the output.

Note that BigDataBench-MT provides two ways of using the above combiner. First, it directly provides some workload replaying scripts, which are the combination results of representative actual workloads (e.g. Hadoop Sort and WordCount) and the Google workload trace. Second, it also supports benchmark users to directly use the above combination technique to match their own data analysis jobs with Google anonymous jobs.

\subsection{Multi-tenant Workload Generator} \label{Section: Multi-tenant Workload Generator}

Based on workload replaying scripts, the workload generator applies a multi-tenant mechanism to generate a mix of workloads using two steps. First, the generator extracts the tenant information from the scripts. For the service and data analysis workloads, this tenant information represents the number of concurrent end users and submitters of analytic jobs, respectively. Second, the generator creates a client for each tenant and emulates the scenarios that a number of end users/job submitters concurrently submit requests/jobs to the system. This multi-tenant framework allows the flexible adjustment of workload scales with guarantee of their realistic mixes. For example, benchmark users can double or halve the size of concurrent tenants, after which the distributions of requests/jobs submitted by these tenants still correspond to those in real workload traces.

%to scale up or down the workload, after which the tenants still generate a realistic mix of workloads.

%and the scaled users
%the scaled workload still follows the mix of real workload traces.

%two advantages.
%the current version
%still follow the realistic mix

%One user

%Another important, but usually ignored factor in traditional benchmarks

\section{Demonstration Description}

\subsection{Chosen Workloads and Workload Traces}

In our demonstration, benchmark users want to evaluate their data center systems using a mix of service and data analysis workloads. The Nutch web search engine \cite{nutchsearch} is used as the example service workload and four Hadoop workloads are used as the example data analysis workloads.
The chosen Hadoop workloads have a variety of workload characteristics: WordCount and Na\"{i}ve Bayes classification are typical CPU-intensive workloads with integer and float point calculations; Sort is the typical I/O-intensive workload and PageIndex is the workload having similar demands for CPU and I/O resources. Both the data generators \cite{ming2014bdgs} and workloads in the demo can be obtained from BigDataBench \cite{opensourceBigDataBench}.

Our demo uses a 24-hour user query logs from Sogou, which include 1,724,264 queries from 519,876 end users, as the basis to generate realistic search engine service; and uses a 24-hour cluster workload trace from Google, which includes 37,842 anonymous jobs from 2,261 job submitters, as the basis to generate realistic Hadoop jobs.

%20 inputs to train regression models

%d

\subsection{System Demonstration}
%Three steps.
%The first part of the demo will be an overview of the near duplicate detection topic in the data management context

%In MapDupReducer's setup dialog, a user has the option to select a data source, set a parameter, and upload a custom stop word list.

\begin{figure*}
\centering
  \includegraphics[scale=0.48]{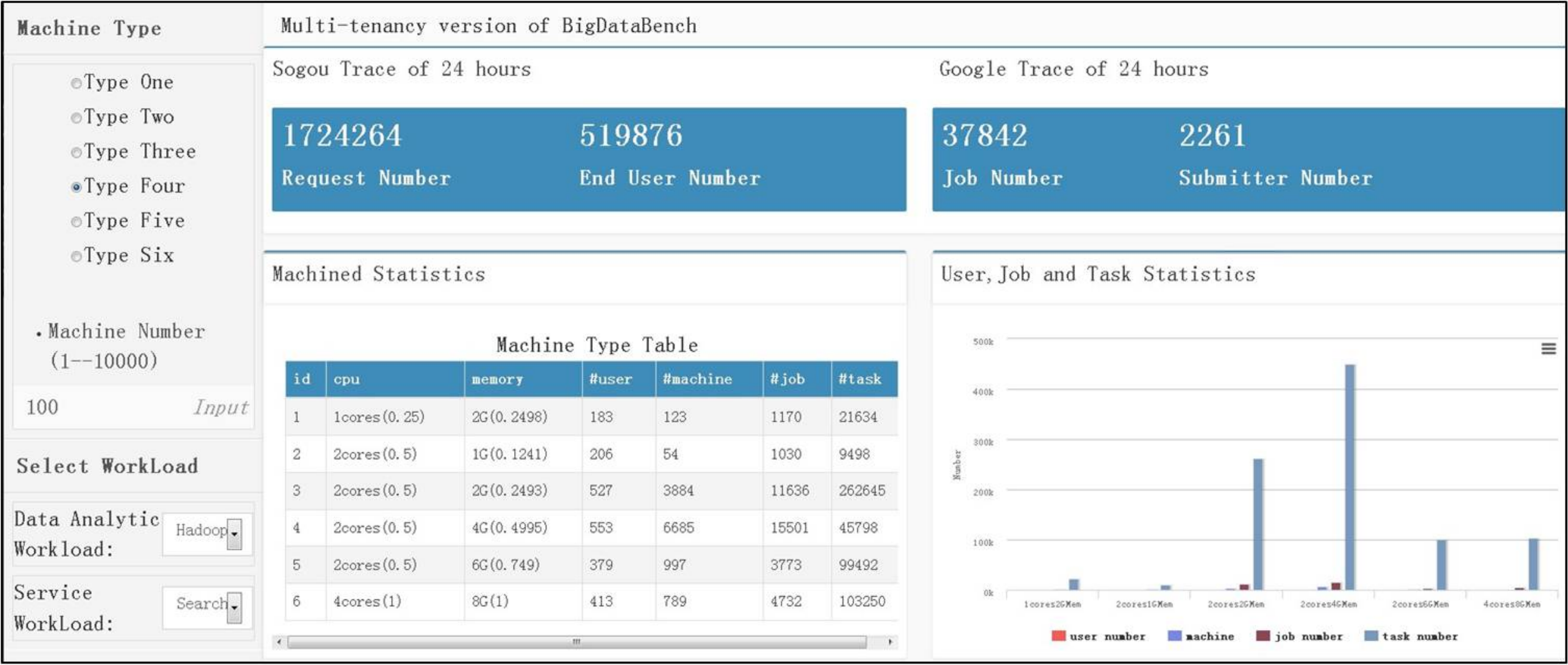}\\
  \caption{System Demonstration Screenshots: Step 1}
  \label{Fig: Demon1}
\end{figure*}

\begin{figure*}
\centering
  \includegraphics[scale=0.48]{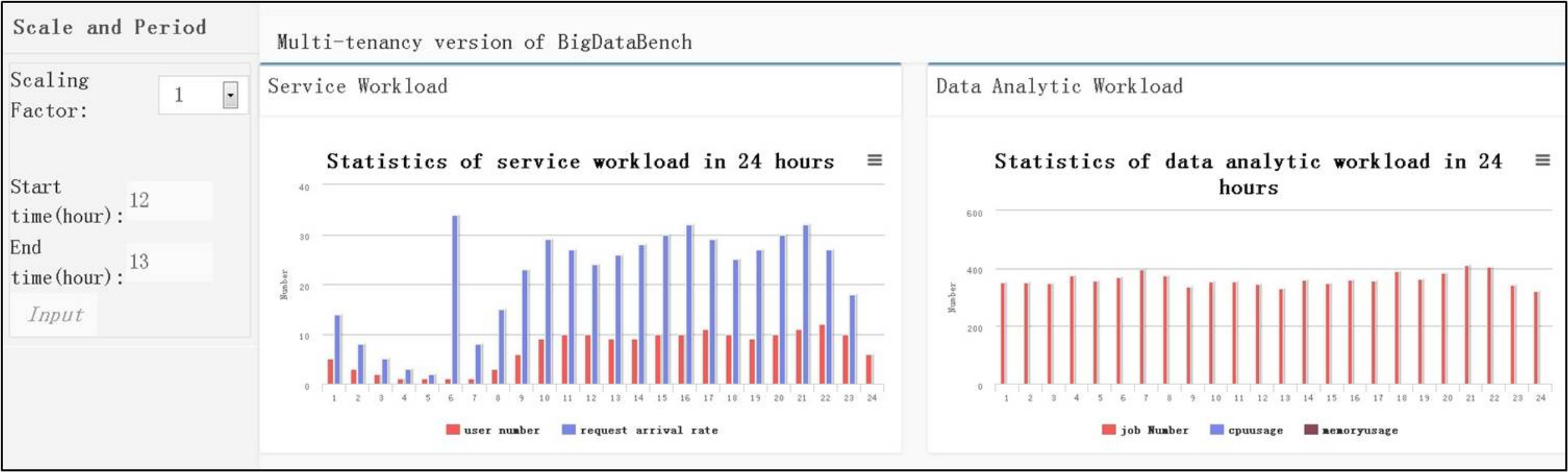}\\
  \caption{System Demonstration Screenshots: Step 2}
  \label{Fig: Demon2}
\end{figure*}

\begin{figure*}
\centering
  \includegraphics[scale=0.48]{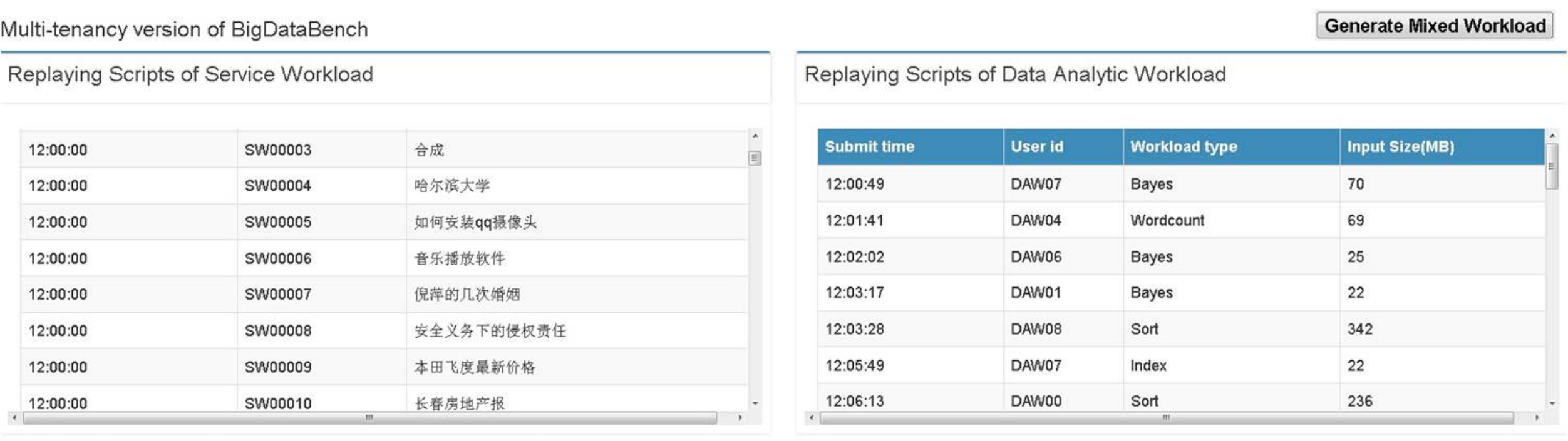}\\
  \caption{System Demonstration Screenshots: Step 3}
  \label{Fig: Demon3}
\end{figure*}

BigDataBench-MT provides a visual interface in the \emph{Benchmark User Portal} to help benchmark users make appropriate benchmarking decisions. This portal provides users necessary information, allows them input benchmarking requirements and executes system evaluations on their behalf. The whole process consists of three steps, as shown in Figures \ref{Fig: Demon1}, \ref{Fig: Demon2} and \ref{Fig: Demon3}, respectively.

\emph{Step 1. Specification of tested machines and workloads}. The first step of the demo presents an overview of workload traces (i.e. Sogou and Google traces) and the data center status, including the six types of machines, their machine number and configurations, and the user, job and task statistics in these machines. This information assists benchmark users to select the type and number of machines to be evaluated, and the workloads they want to use. Suppose users select \emph{Type Four} of the machines with 2 process cores and 4GB memory and 100 machines to be tested, the workload traces belonging to these machines are extracted and forwarded to the next step.

\emph{Step 2. Selection of benchmarking period and scale}. At this step, users have the option to select the period and scale for their benchmarking scenarios.
To facilitate this selection, BigDataBench-MT shows the statistic information of both the service workload (including its number of requests and end users per second) and the data analysis workloads (including their number of jobs and average CPU, memory resource usages) at each of the 24 hours. Suppose users select the benchmarking period of 12:00 to 13:00 and the scale factor is 1 (that is, no scaling is needed). The workload traces belonging to this period are selected for step 3.

\emph{Step 3. Generation of mixed workloads}. After both workloads and traces have been selected, the final step employs the combiner described in Section \ref{Section: Combiner of Workload and Workload Traces} to generate workload replaying scripts for both the service and data analysis workloads, and sends these scripts as feedback to users. In the matching of actual Hadoop jobs with anonymous ones, we tested each Hadoop workload type using 20 different input sizes to build the regression models. Based on the replaying scripts, benchmark users can press the "Generate mixed workload" button to trigger the multi-tenant workload generator, in which each tenant is an independent workload generator and multiple tenants generate a mix of realistic workloads.

%After user specification, the workloads are generated using the multi-tenant workload generator, in which each tenant is a generator of a specific workload provided by BigDataBench.

\section{Future Work}
There are multiple avenues for extending the functionality of our benchmark tool. A first step will be to support more actual workloads. Given that there are 33 actual workloads in the BigDataBench and many workloads (e.g. Bayes classification and WordCount) have three versions of implementations (Hadoop, Spark and MPI), adding more workloads to BigDataBench-MT will be helpful to support wider benchmarking scenarios. We also plan to extend our multi-tenant workload generator to support different classes of tenants and allow users to apply different priority disciplines in workload generation.

\section{Acknowledgements}
This work is supported by the National High Technology Research and Development Program of China (Grant No. 2015AA015308), the National Natural Science Foundation of China (Grant No.61502451), and the Key Technology Research and Development Programs of Guangdong Province, China (Grant No.2015B010108006).

\bibliographystyle{abbrv}
\bibliography{reference}

%\bibliography{reference}

\end{document}